\documentclass{JHEP}
\usepackage{epsfig}
\usepackage{amssymb}
\usepackage[active]{srcltx}

\newcommand{\bea}{\begin{eqnarray}}
\newcommand{\eea}{\end{eqnarray}}
\newcommand{\bean}{\begin{eqnarray*}}
\newcommand{\eean}{\end{eqnarray*}}

\def\O #1{\overline{#1}}
\def\D #1{\dot{#1}}

\def\ba{\begin{array}}
\def\ea{\end{array}}
\def\beq{\begin{equation}}
\def\eeq{\end{equation}}
\def\braket#1{\left\langle #1 \right\rangle}

\def\bbar#1{ \overline #1}

\def\a{{\alpha}}
\def\b{{\beta}}
\def\d{{\rm d}}
\def\da{{\dot{\alpha}}}
\def\db{{\dot{\beta}}}
\def\th{{\theta}}
\def\bth{{\overline{\theta}}}

\preprint{
SNUST 0306002\\
{\tt hep-th/0306215}}
\title{Deformed Superspace, ${\cal N}={1 \over 2}$
Supersymmetry \& (Non)Renormalization
Theorems}
\author{ Ruth Britto ${}^a$, Bo Feng ${}^a$, Soo-Jong Rey ${}^{a,b}$\\
~~~~~~~~\\
${}^a$ Institute for Advanced Study\\
Einstein Drive, Princeton, NJ 08540\\
~~~~~~~~\\
${}^b$ School of Physics \& BK-21 Physics Division\\
Seoul National University, Seoul 151-747 KOREA\\
~~~~~~\\
\email{britto,fengb@ias.edu, sjrey@gravity.snu.ac.kr} }
\abstract{We consider a deformed superspace in which
the coordinates
$\theta$ do
not anticommute, but satisfy a Clifford algebra.  We present
results on the properties of ${\cal N}={1 \over 2}$
supersymmetric theories of chiral superfields in deformed superspace, 
taking the
Wess-Zumino model as the prototype.  We prove new
(non)renormalization theorems: the F-term is radiatively corrected
and becomes indistinguishable from the D-term, while the $\O {\rm
F}$-term is not renormalized. Supersymmetric vacua are critical
points of the antiholomorphic superpotential. The vacuum energy is
zero to all orders in perturbation theory. We illustrate these
results with several examples.}
\begin{document}
%

\section{Introduction}
It has been known that noncommutative geometry arises quite
naturally in string theory \cite{connesdouglasschwartz}: by
turning on a constant Kalb-Ramond $B$-field, Grassmann-even
coordinates $x^m$ are made noncommuting and obeying the Heisenberg
algebra:
\bea \label{C-def}[x^m, x^n ] = i \th^{mn}. \nonumber \eea
In the Seiberg-Witten decoupling limit \cite{seibergwitten},
noncommutative field theories arise on the worldvolume of
D-branes. These field theories have revealed surprising features
reminiscent of closed strings, such as UV-IR mixing \cite{uvir} and
nonlocal observables (open Wilson lines) \cite{owl}, both of which
turn out to be deeply intertwined \cite{summary}. Interplay
between noncommutative field theories and string theories has been
a rich and fruitful source for better understanding of both.

A natural question is whether a noncommutative {\sl
super-geometry} can arise from string theory as well, where,
instead of the Grassmann-even ones, the Grassmann-odd coordinates
$\th^\a (\a = 1,2)$ are made noncommuting and obeying a Clifford
algebra:
\beq \left\{ \th^\a, \th^\b \right\} = C^{\a\b}.
\label{ncrelation} \eeq
Recently, motivated partly by the development of a gauge theory -
matrix model correspondence and its underlying string theory setup
\cite{dijkgraafvafa}, an affirmative answer to the question was
obtained: turning on Ramond-Ramond graviphoton field strength, the
noncommutativity Eq.(\ref{ncrelation}) emerges again quite
naturally from string theories \cite{oogurivafa, stonybrook,early}. Its
consequences are potentially far reaching, and the development
calls for better understanding of field theories defined on
noncommutative superspace Eq.(\ref{ncrelation}). Ground-breaking
work in this direction appeared recently in \cite{seiberg}. 
To
distinguish them from noncommutative supersymmetric field theories
in which Grassmann-even coordinates are noncommuting, we will refer
to the theories under consideration as{ \it deformed supersymmetric
field theories}.

In particular, we would like to consider (\ref{C-def}) as the only
deformation of the superspace; all other algebra on the superspace
remains the same as usual once chiral coordinates are adopted
\cite{seiberg}. This deformation preserves ${\cal N} = {1 \over
2}$ supersymmetry, as the supercharges $Q_\a$, the generators of
$\th^\a$-translation, are conserved (see Eq.(\ref{ncrelation}))
while the $\bbar Q_\da$ are broken explicitly.

In this paper, we study how this deformation would modify quantum
dynamics of supersymmetric field theories, paying particular
attention to consequences of nonlocality in the superspace caused
by Eq.(\ref{ncrelation}). We find quite a few surprises. We will
build our analysis upon both superspace Feynman diagrammatics and
symmetry considerations. For simplicity and clarity of
presentation, we study primarily the deformation of the
Wess-Zumino model. However, because our analysis is
sufficiently general, the final results are equally
valid for other theories. We will report analysis for other field
theories in separate publications \cite{workinprogress}.

Our results are summarized as follows. \hfill\break
$\bullet$ The quantum effective action of the deformed
Wess-Zumino model is expressed as
\beq \Gamma[\Phi, \bbar \Phi] = \sum_{n=0}^\infty {1\over n!}
\int \d^4 x_1 \cdots \d^4 x_n \d^2 \th^2 \d^2 \bth^2
[F_1(x_1,\th,\bth) \cdots F_n(x_n,\th,\bth) G(x_1,\cdots,x_n;\bth \bth)],
\eeq
where $F_1, \cdots, F_n$ are functions involving background fields
$\Phi, \bbar \Phi$, possibly acted upon by $D_\alpha,
\overline{D}_\a$, {\sl and}, most significantly, by $Q_\a$
but not by $\overline{Q}_\da$. The function
$G(x_1, \cdots, x_n;\bth \bth)$ is the result of superspace loop integrals
and is translationally invariant.  Again,
because of the deformation, $(-{1 \over 4}Q^2)$'s may act
inside the loop integrals, and this results effectively in $\bth^2$-dependence. \hfill\break
$\bullet$ The appearance of $Q_\a$'s is the only modification of
the tree-level Lagrangian once the deformation is considered.
Its presence is anticipated by the observation that the star
product
\beq 
\label{C-prod} A(\th) \star B(\th) \equiv A(\th)
\exp \left( -{1 \over 2} C^{ab} \overleftarrow{\partial \over
\partial \th^\a} \overrightarrow{\partial \over \partial \th^\b}
\right) B (\th) \label{originalstar}, 
\eeq
which implements the deformation in terms of the usual anticommuting coordinates,
is re-expressible in terms of the chiral supercharge $Q_\a =
\partial / \partial \th^a$. \hfill\break
$\bullet$ Once $Q_\a$'s are tolerated in the tree-level Lagrangian, one
immediately finds that there is no real distinction between
D-terms and F-terms in the quantum effective action: $(-{1 \over 4}Q^2)$ acting inside loop
integrals generates $\bth^2$-dependence, so the $\d^2 \bth$
integral can yield a F-term. Consequently, once the deformation is
made, not only the D-term but also the F-term is renormalized. In
fact, the two become indistinguishable. \hfill\break
$\bullet$ The antiholomorphic $\O {\rm F}$-term is not renormalized.
This is because, even if  $Q_\a$'s are present, the effective
action $\Gamma (\Phi, \O \Phi)$ cannot yield any $\O {\rm F}$-term by the
$\d^2 \th$ integration.
This ${\cal N}=1/2$
nonrenormalization theorem, which we will prove in this work in
generality, is indeed the prerequisite for the antichiral ring
structure to survive at the quantum level. \hfill\break
$\bullet$ The vacua $|{\rm vac} \rangle, \langle {\rm vac} |$ that
preserve the ${\cal N}={1 \over 2}$ supersymmetry are
characterized by a set of critical points of the antiholomorphic
superpotential, $\O W \, {}'(\O A) = 0$. \hfill\break
$\bullet$ The vacuum energy is not renormalized: the ${\cal N}={1
\over 2}$ supersymmetric vacuum has vanishing energy density and
is stable against radiative corrections. It then follows from the
supersymmetry algebra that
\beq \langle {\rm vac} \vert Q_\a = 0 \qquad {\rm and} \qquad Q_\a
\vert {\rm vac} \rangle = 0. \eeq
This is another prerequisite for existence of the antichiral ring
structure in ${\cal N}={1 \over 2}$ supersymmetry.

This paper is organized as follows. In section 2, we recapitulate
the deformation of supersymmetric field theory and derive
superspace Feynman rules. In section 3, we present the ${\cal
N}={1 \over 2}$ (non)renormalization theorems: (1) the F-term is
renormalized, but the $\O {\rm F}$-term is not; (2) the vacuum
energy is zero to all orders in perturbation theory. We report the
general structure and superspace expression of the quantum
effective action. In section 4, we illustrate this result by
computing several lower-order Feynman diagrams. In section 5, to
illustrate the power of the new (non)renormalization theorems, we
show the vanishing of the vacuum energy explicitly, using various
(pseudo)symmetries. We explain how this vanishing preserves the
antichiral ring. We show that supersymmetric vacua are
critical points of the antiholomorphic superpotential. In section
6, we discuss two intriguing observations concerning the general
structure of these theories as well as phenomenological prospects.

\section{Deformed Supersymmetric Field Theory}
\subsection{Setup}
We begin by  recapitulating relevant aspects of the deformation \cite{seiberg}
of supersymmetric field theories \cite{superspace,wessbagger} and by setting
up the deformed Wess-Zumino model in a form suitable for our
analysis.

Take ${\cal N}=1$ superspace $\cal S$ in {\sl Euclidean} four
dimensions. Choose the `chiral basis' of the superspace
coordinates $z^A = (y^m, \th^\a, \bth^\da)$, where $y^m \equiv
(x^m + i \th \sigma^m \bth)$ are the four real Grassmann-even
coordinates, and $\th^\a, \bth^\da$ are two independent Weyl
Grassmann-odd coordinates. In the coordinates adopted, the
superspace derivatives and supersymmetry generators are given by
\bea
D_\alpha & = & +{\partial \over \partial \th^\alpha}+2 i
\sigma^m_{\alpha \D \alpha} \O \th^{\D \alpha} {\partial
\over \partial y^m}, \qquad \O D_{\D \alpha}=-
{\partial \over \partial \O \th^\alpha}, \label{D-def} \\
\O Q_{\D \alpha} & = &- {\partial \over \partial \O
\th^\alpha}+2i \th^\alpha \sigma^m_{\alpha \D
\alpha}{\partial \over \partial y^m}, \qquad Q_\alpha=
+{\partial \over \partial \th^\alpha}  \label{Q-def}.
\eea
Chiral and antichiral superfields $\Phi, \O \Phi$ are defined by
$\O D_{\D \alpha} \Phi=0$ and $D_\alpha \O \Phi=0$, respectively.
Their expansion in component fields is
\bean
\Phi(y,\th) & = & A(y) + \sqrt 2 \th \psi(y) + \th \th F(y); \\
\bbar \Phi(\bbar y, \bth) & = & \bbar A (\bbar y) + \sqrt 2 \bth \bbar\psi (\bbar y) + \bth \bth \bbar F (\bbar y).
\eean

The deformed theory may then be defined as follows: in the chiral
coordinates adopted, multiply superfields via the $\star$-product
Eq.(\ref{C-prod}). As demonstrated in \cite{seiberg}, the
$\star$-product of chiral superfields is again a chiral
superfield; likewise, the $\star$-product of antichiral
superfields is  again an antichiral superfield. Consequently,
the deformation of a supersymmetric field theory is defined by
replacing all superfield multiplications by $\star$-product
multiplcations. Thus, for the Wess-Zumino model, the deformation
leads to the Lagrangian density
\bea L_{\rm WZ-def} = L_{D} + L_{F} + L_{\O F}, \eea
where
\bea L_{D} &=& \Big[\Phi \star \bbar \Phi \Big]_{\th^2 \bth^2} =
\overline{A} \Box A -i \overline{\psi}
\overline{\sigma}^m \partial_m \psi + \overline{F} F \label{D} \\
L_{F} &=& \left[ {1 \over 2} m \Phi \star \Phi + {g \over 3}
\Phi \star \Phi \star \Phi \right]_{\th^2} \nonumber \\
&=& m \Big(A F -{1 \over 2}\psi \psi \Big) + g \Big(AAF - A\psi
\psi \Big) -{g \over 3} |C| FFF
 \label{F} \\
L_{\O F} &=& \left[ {1 \over 2} \bbar m \bbar \Phi \star
\bbar \Phi + {\bbar g \over 3} \bbar \Phi \star \bbar \Phi \star
\bbar \Phi \right]_{\bth^2} \nonumber \\
&=& \bbar m \Big(\overline{A} \overline{F} - {1 \over 2}
\overline{\psi} \overline{\psi} \Big) + \bbar g \Big(\O A \O A
\overline{F} - \overline{A} \overline{\psi}\overline{\psi} \Big).
\nonumber \eea
We see that the sole effect of the deformation is that the F-term
receives a new contribution proportional to
$|C|$, the determinant of $C^{\a\b}$. Notice that this new term
can be expressed in terms of ordinary products as \footnote{Recall
that $Q^2 \Psi$ is also a (anti)chiral superfield if $\Psi$ is so.
This follows trivially from the fact $\{ Q, D \} = \{Q, \O D\} =
0$.}
\bea \Delta_{\rm def} L_F = -{g \over 3} |C| \Big[ \Big((-{1 \over 4}Q^2)
\Phi \Big)^2 \Phi \Big]_{\th^2}. \label{cterm} \eea
Consequently, one can view the deformed Wess-Zumino model as the
ordinary Wess-Zumino model, where superfield multiplication is
standard, with a new addition Eq.(\ref{cterm}) to the F-term. That
is, the deformed Wess-Zumino model is definable by the
Lagrangian
\bea L_{\rm WZ-def} = \Big[\Phi \O \Phi \Big]_{\th^2 \bth^2} + \Big[
{m \over 2}\Phi \Phi + {g \over 3} \Phi \Phi \Phi \Big]_{\th^2} +
\Big[ {\O m \over 2} \O \Phi \O \Phi + {\O g \over 3} \O \Phi \O
\Phi \O \Phi \Big]_{\bth^2} + \Delta_{\rm def} L_F. \label{lag1}\eea
In components, the Lagrangian is
\bea L_{\rm WZ-def} &=& \O A \Box A - i \O \psi \O \sigma^m
\partial_m \psi + \O F F + m \left( A F - {1 \over 2} \psi \psi
\right) + \O m \left(\O A \O F - {1 \over 2} \O \psi \O \psi
\right) \nonumber \\
&+& g (A A F - A \psi \psi) + \O g ( \O A \O A \O F - \O A \O \psi
\O \psi) - {g \over 3} |C| FFF . \label{lag2} \eea

Adding the last term in Eqs.(\ref{lag1}, \ref{lag2}) renders the
theory quite novel. In ordinary Wess-Zumino model, because of
${\cal N}=1$ supersymmetry, one is not allowed to introduce $Q_\a$
and $\O Q_\da$ in the Lagrangian. With the deformation, however,
the $\O Q_\da$-part of the supersymmetry is broken explicitly, so
here $Q_\a$'s are allowed. Conversely, if $ Q_\a$'s are present in
the Lagrangian, the antichiral part of the ${\cal N}=1$ supersymmetry
(associated with translation of $\bth$) is broken, and the theory
preserves ${\cal N}={1 \over 2}$ supersymmetry only. In fact, as
mentioned above, the definition of $\star$-product is expressible
entirely in terms of $Q_\a$'s
\bea \Phi_1 (y, \th) \star \Phi_2 (y, \th) = \Phi_1 (y,
\th) \exp \left( -{1 \over 2} C^{\a\b}
\overleftarrow{Q_\a}\overrightarrow{Q_\b} \right) \Phi_2 (y,
\th) \label{Qstar} \eea
at fixed $y$ in the chiral coordinates. Therefore, it is not
surprising that, when recast in ordinary superspace formulation,
the chiral supercharges $Q_\a$ show up in the Lagrangian.
Consequences of such explicit ${\cal N}={1 \over 2}$ supersymmetry
breaking are quite interesting, as we demonstrate in detail in later
sections.
\subsection{Deformed Feynman rules}
We now present Feynman rules for the deformed Wess-Zumino model.
They are derived by straightforward application of the standard
method \cite{superspace, wessbagger, BK}. We present the rules both in
component form and in superfield form. We also summarize various
identities utilized for later computations.

The superspace Feynman rules are:
\begin{itemize}
\item (1) Use the so-called GRS propagators \cite{GRS} for
internal lines. They are given as follows:
\bea \label{super-pro} \left< \Phi(z) \O \Phi(z')\right> & = &
{i\over \Box-m \overline{m}} \delta^8(z-z'),\\
\left< \Phi(z) \Phi(z') \right> & = & {- i \O m  \over (\Box-m \O
m)}
{1 \over \Box} \left(- {1 \over 4}D_z^2 \right) \delta^8(z-z'), \nonumber \\
\left< \O \Phi(z) \O \Phi(z') \right> & = &{-i  m \over (\Box-m \O
m)} {1 \over \Box} \left( - {1  \over 4}\O D_z^2 \right)
\delta^8(z-z'),\nonumber \eea
where $\delta^8(z-z') = \delta^4(x-x')
\delta^2(\th-\th')\delta^2(\O\th-\O \th')$.
For later reference, we also record the propagators in component
form
\bea \label{pro-component} \langle A(x)\overline{A}(x')\rangle & =
& {i\over \Box-m \overline{m}} \delta^4(x-x') \qquad \qquad
\langle F(x)\overline{F}(x')\rangle =
{i\, \Box\over \Box-m \overline{m}}\delta^4(x-x') \\
\langle A(x) F(x') \rangle & = &{-i \, \overline{m}\over \Box-m
\overline{m}}\delta^4(x-x') \qquad \qquad
\langle\overline{A}(x)\overline{F}(x')\rangle  =
{-i \, m \over \Box-m \overline{m}}\delta^4(x-x') \nonumber  \\
\langle \psi_\alpha(x) \psi^\beta (x')\rangle & = &
\delta_\alpha^\beta {i \, \overline{m}\over \Box-m
\overline{m}}\delta^4(x-x') \quad \, \quad \langle
\overline{\psi}^{\dot{\alpha}}(x)\overline{\psi}_{\dot{\beta}}
(x')\rangle  =  \delta^{\dot{\alpha}}_{\dot{\beta}}
{i \, m \over \Box-m \overline{m}}\delta^4(x-x') \nonumber \\
&& \qquad \langle \psi_\alpha(x)
\overline{\psi}_{\dot{\beta}}(x')\rangle =
{\sigma^m_{\alpha\dot{\beta}}\partial^x_{m} \over \Box-m
\overline{m}}\delta^4(x-x'), \nonumber \eea where
$\Box=\partial_m\partial_m$.
\item (2) There are two chiral vertices, $\Phi^3$, $(-{1 \over 4}Q^2 \Phi)^2
\Phi$; and one antichiral vertex, $\O \Phi^3$. Associated with
every chiral vertex carrying $n$ internal lines, $(n-1)$ factors
of $\left(-{1 \over 4}\O D^2 \right)$ act on some arbitrary
$(n-1)$ internal lines. Likewise, associated with every
antichiral vertex carrying $n$ internal lines, $(n-1)$ factors of
$\left(-{1 \over 4}D^2 \right)$ act on some arbitrary $(n-1)$
internal lines. All external lines arise without any of these
factors.
\item (3) For the new vertex $\left(-{1 \over 4}Q^2
\Phi \right)^2 \Phi$, two factors of $\left( -{1 \over 4}Q^2 \right)$ are attached to an arbitrary two of the three (external or
internal) lines.
\item (4) Associated to every vertex, multiply the appropriate factor of ${1
\over 3}g, {1\over 3} \O g$ or $-{1 \over 3} g |C|$, and perform
the superspace integral $\int \d^8 z$.
\item (5) Compute the standard combinatoric factors for a given
theory.
\end{itemize}
Notice that only rule (3) is new for the deformed theory. As
such, the deformed Feynman rules are quite general, and extend
straightforwardly to other field theories.

Recall that, for a chiral superfield $\Phi$, $\Xi \equiv (-{1
\over 4} Q^2) \Phi$ is also a chiral superfield. Therefore, the
deformed superspace Feynman rules are exactly the same as the
ordinary superspace Feynman rules if we treat $\Xi$ as an
independent chiral superfield and Wick-contract $\Xi$ with the
propagators
\bea \langle \Xi (z) \Phi(z') \rangle &=& \Big( -{Q^2 \over 4}
\Big) \langle \Phi(z) \Phi(z') \rangle \nonumber \\
\langle \Xi (z) \bbar \Phi(z') \rangle &=& \Big( -{Q^2 \over 4}
 \Big) \langle \Phi(z) \bbar \Phi(z') \rangle, \nonumber \eea
and similarly for antichiral counterparts.

\section{(Non)-Renormalization Theorems}
We now compute the one-particle-irreducible effective action. We
will keep both $m, \O m$ nonzero so that the effective action is
well defined in the infrared.

Before presenting the general structure of the effective action in
the deformed theory, it will be useful to recollect the standard
non-renormalization theorems of supersymmetric field theories as summarized for example in \cite{BK}:
{\it
\begin{itemize}
\item {\tt Theorem 1:}~~Each term in the effective action is
expressible as a superspace integral over a single $\d^2 \th
\d^2 \bth$.
\item {\tt Theorem 2:}~~The general structure of the effective
action is given as
\bea \Gamma[\Phi, \O \Phi]=\sum_n \int \prod_{j=1}^n \d^4 x_j \int
\d^2 \th \d^2 \bth \, G_n(x_1,...,x_n) F_1(x_1,\th)...
F_n(x_n,\th).
\nonumber
\eea
where $G_n(x_1,...,x_n)$ are translation-invariant functions on
Grassmann-even coordinates and $F(x,\th, \bth)$ are {\sl local}
operators of $\Phi,\O \Phi$ and their covariant derivatives:
\bea F(x,\th, \bth)=F(\Phi, \O \Phi, D\Phi, \O D \O \Phi,...)
\nonumber \eea
\end{itemize}
}
The above theorems, especially Theorem 2, lead immediately to the
following results: (1) energy density of supersymmetric vacuum is
zero because, in this case, there are no $F(x,\th, \bth)$ field
insertions in the effective action, so the $\int \d^2 \th \d^2
\bth$ integral gives zero; (2) the holomorphic and antiholomorphic
parts are not renormalized. The reason is that to get holomorphic
part one needs to integrate out $\int \d^2 \O \th$. However, as
there is no $\Box^{-1}$ in the effective action, one cannot do
that by combining it with the $D^2$ operator. A similar argument
holds for the antiholomorphic part.

Now in our deformed theory, Theorem 1 is not modified. The
proof goes exactly the same as the ordinary theory. However, Theorem 2 is modified crucially by the new vertex with insertion
of the operator $-{1 \over 4}Q^2$.  It follows from the Feynman rules in the previous section that this operator affects loop integrals in a way similar to $-{1 \over 4}D^2$ and  $-{1 \over 4}\O D^2$.  Thus we derive the following new theorem.
{\it
\begin{itemize}
\item {\tt Theorem 2} {\rm [after deformation]:}~~ The general
structure of the effective action is given as
\bea \Gamma[\Phi, \O \Phi]=\sum_n \int \prod_{j=1}^n \d^4 x_j \int
\d^2 \th \d^2 \bth \, G_n(x_1,...,x_n; \bth\bth)
F_1(x_1,\th, \bth)... F_n(x_n,\th, \bth), \eea
where $G_n(x_1,...,x_n; \bth\bth)$ are translation-invariant
functions on Grassmann-even coordinates {\bf and} possible
insertion of $\O \th \O \th$, while $F(x,\th, \bth)$ are
local operators of $\Phi,\O \Phi$, their covariant derivatives, {\bf
and} the action of the chiral supercharge $Q$:
\bea F(x,\th, \bth)=F(\Phi, \O \Phi, D\Phi, \O D \O \Phi,
Q\Phi, Q\O \Phi...) \nonumber\,  . \eea
\end{itemize}
}
We will present later explicit computations and symmetry arguments to substantiate the general structure of the effective action as claimed, but the crux of the new theorem stems from insertion of $Q_\alpha$ and its effects.

Using the modified Theorem 2, we are now able to derive the
following results: (1) energy density of supersymmetric vacuum is
still zero. Although the $\O\th \O \th$-dependence $G(x_1, \cdots,
x_n; \bth\bth)$ would be able to render the $\int \d^2 \bth$
integral nonzero, in the absence of any $F(x, \th, \bth)$
insertions, the $\int \d^2 \th$ integral still vanishes; (2) The
antiholomorphic part is still not renormalized, because the $\int
\d^2 \th$ are not absorbable, for the same reason as in the
ordinary Wess-Zumino model; (3) However, the holomorphic part {\it
is renormalized}. The reason is that now we have the $\O\th \O
\th$ insertion from $G(x_1, \cdots, x_n; \bth\bth)$, which can
absorb the $\int \d^2 \O\th$ integral. Because of this, the
D-terms with pure chiral fields and holomorphic F-terms are not
distinguishable, and in fact both D-terms and F-terms are unified
in ${\cal N}={1 \over 2}$ supersymmetry. We emphasize that this is
the feature that was not evident from the classical consideration
\cite{seiberg}, but was revealed only after full quantum effects
are taken into account.

\section{Illustration by Diagrams}

In this section we outline the computation of a few Feynman diagrams that contribute to the effective action with a single factor of $|C|.$
These examples illustrate the diagrammatic consequences of the new vertex, $(-{1 \over 4} Q^2 \Phi)^2 \Phi.$   We will see how the extra factors of $Q^2$ can appear sometimes on the external lines, sometimes as factors of $\bbar \th \bbar \th$, and sometimes disappear altogether.

We begin with the simplest such diagram, $\braket{FF}$ at one loop, which generates the term $\bbar \th \bbar \th (Q^2 \Phi) \Phi$ in the effective action.  The next example is similar, a one-loop diagram for $\braket {{\bbar A}FF}$ that generates
$\bbar \th \bbar \th (Q^2 \Phi) \Phi \bbar \Phi$.  The last few examples show how to generate terms without $\bth \bth$ or $Q^2$.  We also discover nonholomorphic corrections to the couplings of seemingly holomorphic F-terms.

The principal tools in these calculations are integration by parts
and the identities listed below for differential operators acting
on superspace delta functions.

\bea \label{identity1}
\epsilon^{\alpha \beta} (\sigma_{\alpha
\dot{\alpha}}^m \overline{\th}^{\dot{\alpha}}
\partial_{m})(\sigma_{\beta \dot{\beta}}^n \overline{\th}^{\dot{\beta}}
\partial_{n}) = -\overline{\th}\overline{\th} \Box
&\qquad&
(\th^\alpha\sigma_{\alpha \dot{\alpha}}^m
\partial_{m})(\th^\beta  \sigma_{\beta
\dot{\beta}}^n\partial_{n}) \epsilon^{ \dot{\alpha}\dot{\beta}}  =
+ \th \th\Box
\eea
\bea \label{identity2}
 ({-D^2 \over 4})  ({-\O D^2 \over 4})   D^2 = D^2 \Box
&\qquad& D^2 D^{'2}\delta^8(z-z') = D^{'2}D^2\delta^8(z-z')
\nonumber
\\
 ( {-\O D^2 \over 4} ) ( {-D^2 \over 4}  ) \O D^2 = \O D^2 \Box
&\qquad& D^2 \O D^{'2}\delta^8(z-z')  =  \O D^{'2}D^2\delta^8(z-z')
\\
({-D^2 \over 4} ) ({-\O D^2 \over 4} ) \delta^4(\th-\th')\Big|_{\th=\th'} = 1
&\qquad& D_\alpha \delta^8(z-z') = -D'_\alpha\delta^8(z-z')
\nonumber
\eea
\bea \label{identity3}
({-Q^2 \over 4} ) ( {-D^2 \over 4}  ) = \O\th
\O\th\Box ({1\over 4} \epsilon^{\alpha\beta} {\partial \over
\partial \th^\alpha} {\partial \over \partial
\th^\beta})
&\qquad&
( {-Q^2 \over 4} ) \Phi(y,\th) = F(y)
\\
({-Q^2 \over 4} )({-\O D^2 \over 4} )\delta^4(\th-\th')|_{\th=\th'}
= 1
&\qquad&
({-Q^2 \over 4} )({-D^2 \over 4}) ({-\O D^2 \over 4} ) \delta^4(\th-\th')|_{\th=\th'} = \O\th\O\th \Box
\nonumber
\eea

\subsection{A term with $Q^2$: two-point function}\label{2ptsec}

\EPSFIGURE[ht]{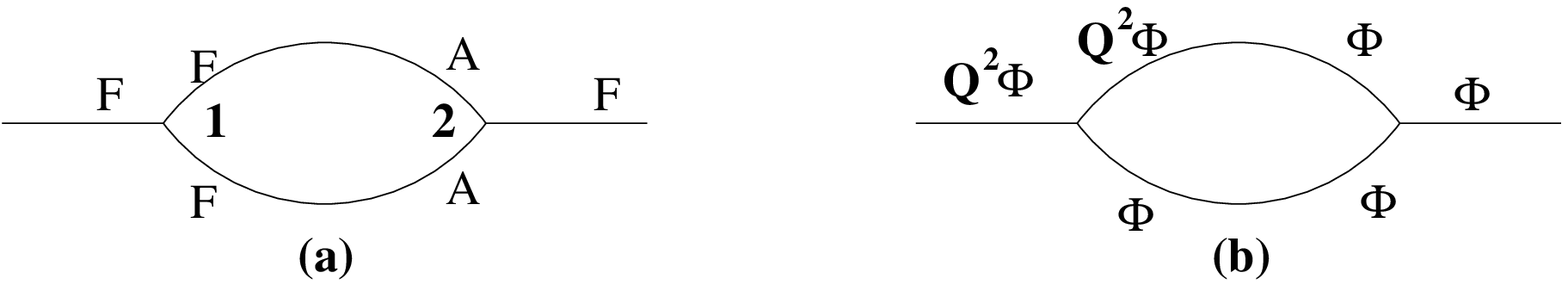,width=14cm} {The one loop correction
to $\langle FF \rangle$. (a) the component diagram (b) the
corresponding superfield diagram. \label{f:twopoint} }

The $\langle FF \rangle$ correction at one loop is the simplest
new contribution to the effective action with $Q^2$ acting on the
external line. See Figure \ref{f:twopoint}.  The term in the
effective action is of the form $\bbar \th \bbar \th (Q^2 \Phi)
\Phi$.

In component form, we have

\bea\label{comp2}
 & & 6(-{g\over 3} |C|) g \int \d^4x_1   \d^4x_2 F(x_1) F(x_2)
\left[ {-i \overline{m}\over \Box-m \overline{m}} \delta(x_1-x_2)
\right]^2  \nonumber \\
& = & 2 g^2 |C| \O m^2\int \d^4x_1   \d^4x_2 F(x_1) F(x_2)
\left[ {1\over \Box-m \overline{m}} \delta(x_1-x_2)
\right]^2,
\eea
where 6 is the symmetry factor.  The integral diverges logarithmically.

In superfield form, we have
\bean
& & 18 (-{g\over 3}  |C|){g\over 3} \int \d^4x_1 \d^4x_2
\d^4 \th_1 \d^4 \th_2 ({-Q_1^2\over 4} \Phi(x_1,\th_1))
\Phi(x_2,\th_2)  \\
& &  \left[{-Q_1^2\over 4} {-\O D_1^2\over 4}
{i \O m  D_1^2 \over 4 \Box(\Box-m \O m)} {-\O D_2^2\over 4}
\delta^8(z_1-z_2) \right]
\left[ {i \O m  D_1^2 \over 4 \Box(\Box-m \O m)}\delta^8(z_1-z_2)
\right],\\
\eean
where 18 is the symmetry factor.

The calculation in superfield form proceeds as follows.  In the
first bracketed factor, replace ${-\O D_2^2\over 4}$ by ${-\O
D_1^2\over 4}$ since it is acting directly on the delta function.
This is a manipulation we shall use often.  It is valid because
all preceding operators can be stripped by integration by parts,
the substitution made by the last identity of
Eq.(\ref{identity2}), and then all preceding operators replaced.

Then apply another identity from Eq.(\ref{identity2}) to this
first bracketed factor, to replace the product ${-\O D_1^2\over 4}
{ \O D_1^2 \over 4 \Box} {-\O D_1^2\over 4}$ by ${-\O D_1^2\over
4}$. Integrate by parts under ${-Q_1^2\over 4}$ and ${-\O
D_1^2\over 4}$, noting that ${-Q_1^2\over 4} \Phi$ is a chiral
field, to get \bean & & -2 g^2  |C| \int \d^4x_1 \d^4x_2 \d^4
\th_1 \d^4 \th_2 ({-Q_1^2\over 4} \Phi(x_1,\th_1))
\Phi(x_2,\th_2) \\
& & \left[{i \O m \over 4 (\Box-m \O m)}
\delta^8(z_1-z_2) \right]
\left[{-Q_1^2\over 4} {i \O D_1^2 \O m  D_1^2 \over 4 \Box(\Box-m \O m)}\delta^8(z_1-z_2)
\right].\\
\eean

Finally, use the symmetry properties of delta functions, and
commutation of the operators, to change the indices of the
differential operators in the last factor all to 2.  Note that
this comes at the cost of exchanging the order of $\O D^2$ and
$D^2$, since we can only exercise this symmetry on the operator
directly in front of the delta function.  Then apply
Eq.(\ref{identity3}) to find
\bean & & 2 g^2  |C| \O m^2 \int \d^4x_1 \d^4x_2 \d^4 \th_1
({-Q_1^2\over 4} \Phi(x_1,\th_1)) \Phi(x_2,\th_1) (\O \th_1 \O
\th_1)  \left[{1\over\Box-m \O m} \delta^4(x_1-x_2) \right]^2,
\eean
 which reproduces the expression Eq.(\ref{comp2}) in component
fields.

\subsection{A term with $Q^2$: three-point function}

There are two superfield diagrams at one loop with the new
vertex $\left(-{1 \over 4}Q^2
\Phi \right)^2 \Phi$ 
that generate three-point functions with $Q^2$ acting on
external lines. One of these combines two $\Phi^3$ vertices and
one $(Q^2 \Phi)^2 \Phi$ vertex, generating a term $\bth \bth (Q^2
\Phi) \Phi^2$ in the effective action.  The other, shown in Figure
\ref{f:threepoint}, generates a term $\bbar \th \bbar \th (Q^2
\Phi) \Phi \bbar \Phi$ in the effective action. The calculations
are very similar; here we present the second. Note that while the
contribution from the two-point function in the previous
subsection could have passed for an F-term, this one cannot,
because it includes an antichiral field.

\EPSFIGURE[ht]{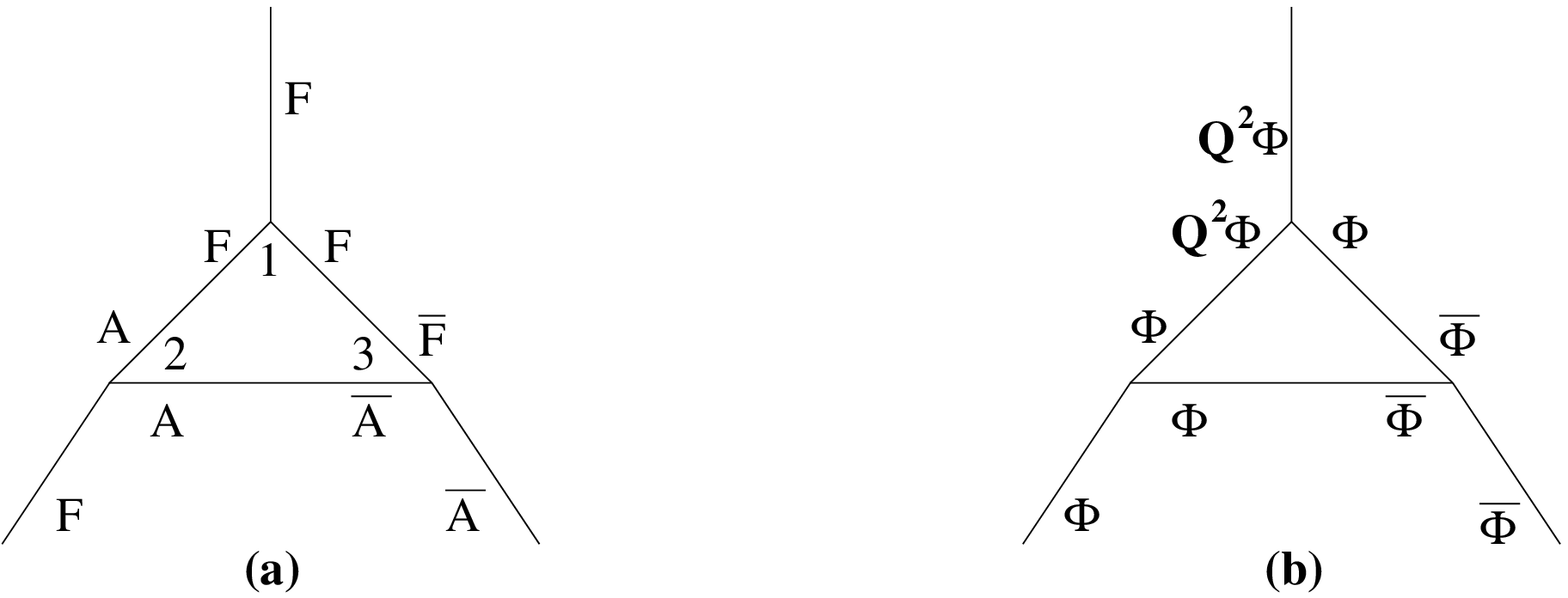,width=12cm}
{The one loop correction to $\langle FF\overline{A}\rangle $.
 (a) the component
diagram (b) the corresponding superfield diagram.
\label{f:threepoint} }

The corresponding diagram in component fields is shown in Figure \ref{f:threepoint}.  The integral form is
\bea\label{thrpocomp}
& & 3 (2^3) (-{g\over 3} |C|) g\O g \int \d^4 x_1 \d^4 x_2 \d^4 x_3
 F(x_1) F(x_2) \O A(x_3) \nonumber \\
& & \left[ {-i \O m \over \Box-m \O m}\delta^4(x_1-x_2)\right]
\left[ {i \Box \over \Box-m \O m}\delta^4(x_1-x_3)\right]
\left[ {i \over \Box-m \O m}\delta^4(x_2-x_3)\right] \nonumber\\
& = & -8 i g^2 \O g  |C|\O m\int \d^4 x_1 \d^4 x_2 \d^4 x_3
 F(x_1) F(x_2) \O A(x_3) \nonumber \\
& & \left[ {1 \over \Box-m \O m}\delta^4(x_1-x_2)\right]
\left[ { \Box \over \Box-m \O m}\delta^4(x_1-x_3)\right]
\left[ {1 \over \Box-m \O m}\delta^4(x_2-x_3)\right].
\eea
Again the contribution diverges logarithmically.

The superfield calculation is much like the previous one.  The expression is
\bean
& & 6^3 (-{g\over 3}|C|) {g\over 3}{\O g\over 3}
\int \d^4 x_1 \d^4 x_2 \d^4 x_3 \d^4\th_1  \d^4\th_2 \d^4\th_3
({-Q_1^2\over 4} \Phi(x_1,\th_1))\Phi(x_2,\th_2)
\O \Phi(x_3,\th_3)\\
& & \left[{-Q_1^2\over 4}  {-\O D_1^2\over 4}
{i \O m  D_1^2 \over 4 \Box(\Box-m \O m)} {-\O D_2^2\over 4}
\delta^8(z_1-z_2) \right]
\left[ {- D_3^2\over 4}  { i\over\Box-m \O m} \delta^8(z_1-z_3) \right]
\left[ { i\over\Box-m \O m} \delta^8(z_2-z_3) \right]. \\
\eean
Upon application of integration by parts and the identities
in Eqs.(\ref{identity2} -- \ref{identity3}), it reduces to
\bea
& &  -8 i g^2 \O g  |C|\O m \int \d^4 x_1 \d^4 x_2 \d^4 x_3 \d^4\th
(\O\th\O\th)
({-Q_1^2\over 4} \Phi(x_1,\th))\Phi(x_2,\th)
\O \Phi(x_3,\th) \nonumber \\
& & \left[{1\over \Box-m \O m}\delta^8(z_1-z_2) \right]
\left[{\Box \over\Box-m \O m}\delta^4(x_1-x_3)
  \right]\left[ { 1\over\Box-m \O m} \delta^4(x_2-x_3) \right],
\nonumber \eea
which is the same result as Eq.(\ref{thrpocomp}).

\subsection{A term in which $Q^2$'s disappear}

There is a two-loop tadpole for $F$, shown in Figure \ref{f:tadpole}, that corresponds to a $\bbar \th \bbar \th \Phi$ term in the effective action.  As shown in the figure, there are three contributions at this order (or two in terms of superfields), but supersymmetry cancels two of them (or one in terms of superfields).
We sketch the computation to observe how the factors of $Q^2$ disappear in the result.

\EPSFIGURE[ht]{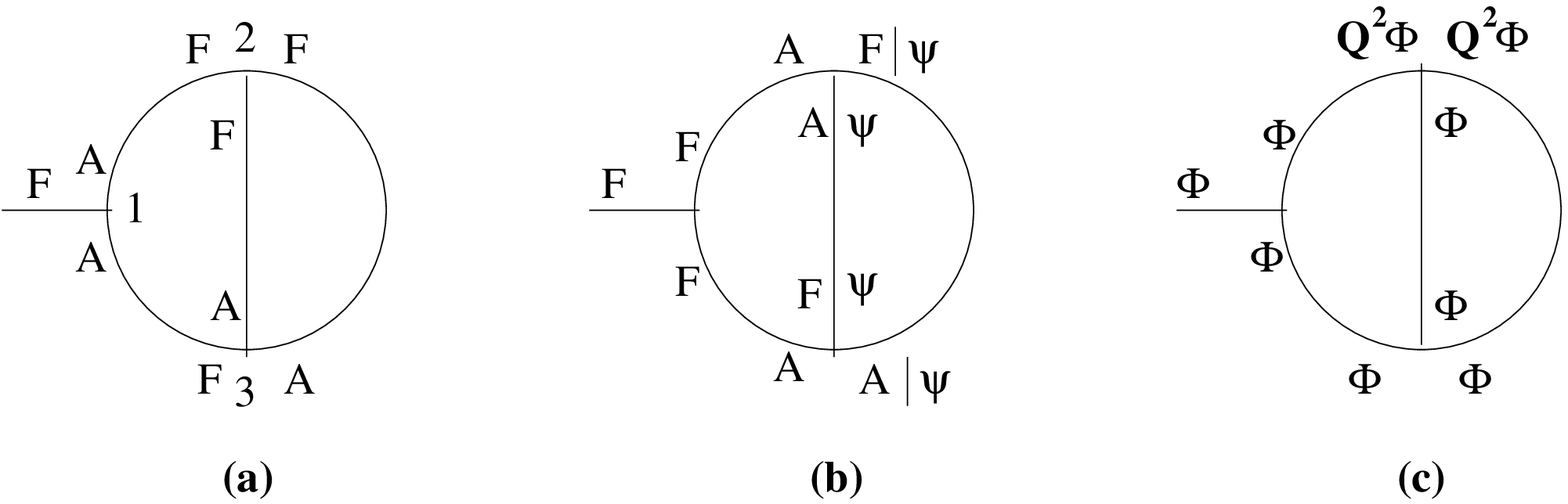,width=14cm} {The tadpole diagram
$\braket F$. There are two distinguished cases: (a) all three
lines of $F^3$ vertex are contracted; (b) one line of  $F^3$
vertex is not contracted. However, for case (b),  bosonic and
fermionic loop contributions cancel each other. (c) is the
corresponding superfield diagram of case (a). \label{f:tadpole} }

The integral in component fields is
\bea\label{tadcomp}
& & 24 (-{g\over 3}|C|) {1\over 2} g^2
\int \d^4 x_1 \d^4 x_2 \d^4 x_3 F(x_1)
\left[{-i\O m\over \Box-m \O m}\delta^4(x_1-x_2)\right] \nonumber \\
& & \left[{-i\O m\over \Box-m \O m}\delta^4(x_1-x_3)\right]
\left[{-i\O m\over \Box-m \O m}\delta^4(x_2-x_3)\right]
\left[{-i\O m\over \Box-m \O m}\delta^4(x_2-x_3)\right] \nonumber\\
& = & -4 g^3 |C| {\O m}^4 \int \d^4 x_1 \d^4 x_2 \d^4 x_3 F(x_1)
\left[{1 \over \Box-m \O m}\delta^4(x_1-x_2)\right]  \nonumber\\
& & \left[{1 \over \Box-m \O m}\delta^4(x_1-x_3)\right] \left[{1
\over \Box-m \O m}\delta^4(x_2-x_3)\right] \left[{1 \over \Box-m
\O m}\delta^4(x_2-x_3)\right]. \eea
The superfield integral is
\bean & & 72 (-{g\over 3}|C|) {1\over 2}({g \over 3})^2 \int
\prod_{a=1}^3 \d^4 x_a \d^4\th_a \Big[{-\O D_3^2\over 4}{i \O m
D_1^2 \over 4 \Box(\Box-m \O m)} \delta^8(z_3-z_1) \Big] \Big[{i
\O m D_2^2 \over 4 \Box(\Box-m \O m)}\delta^8(z_2-z_3)
\Big] \\
&&\left[{-Q_2^2\over 4} {-\O D_2^2\over 4} {i \O m  D_2^2 \over 4
\Box(\Box-m \O m)}{-\O D_1^2\over 4} \delta^8(z_2-z_1)
\right]
\left[{-Q_2^2\over 4} {-\O D_2^2\over 4} {i \O m  D_2^2
\over 4 \Box(\Box-m \O m)}{-\O D_3^2\over 4} \delta^8(z_2-z_3)
\right]. \eean
Manipulations similar to those in the previous
calculations bring the expression into the form
\bean & & -4 g^3 |C| {\O m}^4 \int \prod_{a=1}^3 \d^4 x_a
\d^4\th_a \Phi(x_1,\th_1) \left[{1 \over \Box - m \O m}
\delta^8(z_2-z_1)\right]\left[{1 \over \Box - m \O m}
\delta^8(z_3-z_1)\right] \\
& & \qquad \left[{-Q_2^2 \over 4}{-D_2^2 \over 4}{-\bbar D_2^2
\over 4 \Box}{1 \over \Box - m \O m} \delta^8(z_2-z_3)\right]
\left[{-Q_2^2 \over 4}{-\bbar D_2^2 \over 4}{1 \over \Box - m \O
m} \delta^8(z_2-z_3)\right]. \eean
Now factor the fermionic delta functions separately so that we can
apply Eq.(\ref{identity3}) to replace the actions of the
differential operators in the last two bracketed factors by
$\bth_2^2 \Box$ and $1,$ respectively.  Then we find that the
expression becomes
\bean & &-4 g^3 |C| \O m^4 \int \d^4 x_1 \d^4 x_2 \d^4 x_3 \d^4\th
(\bbar \th \bbar \th) \Phi(x_1,\th)
\left[{1 \over \Box - m \O m} \delta^8(x_2-x_1)\right] \nonumber \\
& &\left[{1 \over \Box - m \O m} \delta^8(x_3-x_1)\right] \left[{1
\over \Box - m \O m} \delta^8(x_2-x_3)\right] \left[{1 \over \Box
- m \O m} \delta^8(x_2-x_3)\right], \eean
 which confirms Eq.(\ref{tadcomp}).

\subsection{A term without $\bbar \th \bbar \th$}

We also obtain terms in the effective action with $Q^2$ but
without $\bbar \th \bbar \th$. An example is shown in part (a) of
 Figure \ref{f:FFgbar},
 corresponding to the term $(Q^2 \Phi)^2 \Phi \bbar \Phi$ from e.g.~the four-point function $\braket{F^3{\bbar F}}$.
 In this particular calculation, both factors of $Q^2$ end up on
 external lines after integration by parts.  This contribution again diverges logarithmically.



\EPSFIGURE[ht]{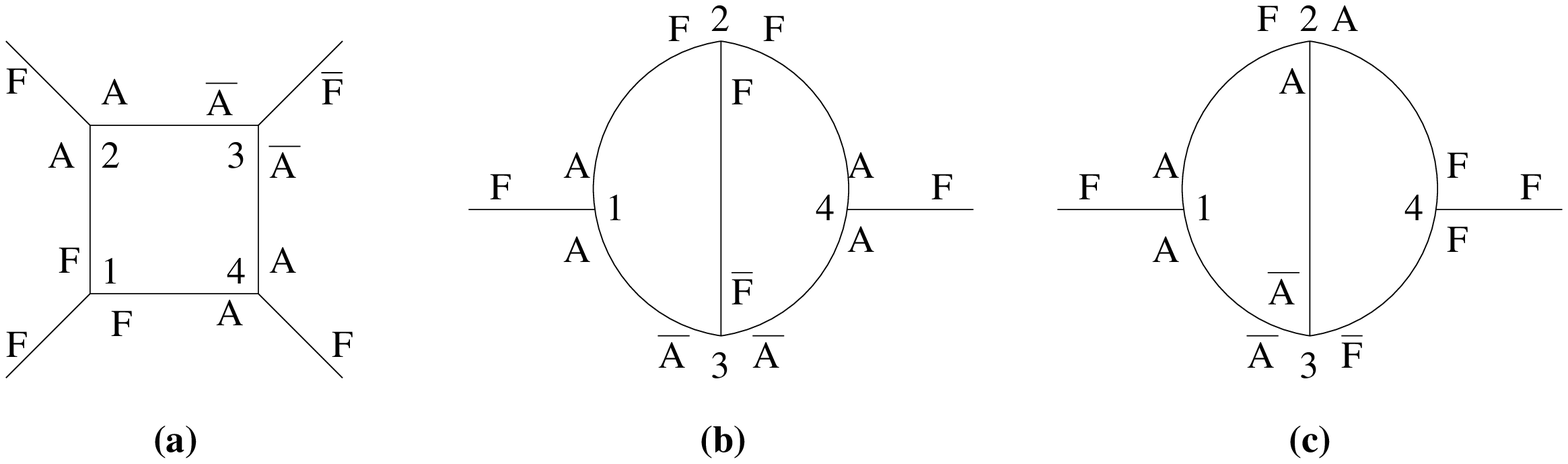,width=14cm} { (a) The diagram without
the generation of $\O \th \O \th$ in the calculation of superfield
computation. (b,c) The diagrams of a holomorphic part which depend
on the antiholomorphic parameter $\O g$. \label{f:FFgbar} }

\subsection{`F-terms' are not holomorphic}
As implied by the new (non)renormalization theorem, the F-term is
no longer holomorphic. The second example (parts (b,c) of Figure
\ref{f:FFgbar}) shows a logarithmically divergent two-loop
function  contribution to $\langle FF \rangle$ that yields the
term $\bbar \th \bbar \th (Q^2 \Phi) \Phi$ in the effective
action, as in subsection \ref{2ptsec}. Because this contribution
involves the antiholomorphic coupling, $\bbar g,$ we see that
terms that can {\em appear} like F-terms are not restricted to
involve only holomorphic couplings. We can no longer distinguish
F-terms and D-terms.


\subsection{A term with neither $Q^2$ nor $\bbar \th \bbar \th$}

Finally, the three-loop example drawn in Figure \ref{f:threeloop}
 shows that we can generate terms with
neither $Q^2$ nor $\bbar \th \bbar \th$.  This is a contribution
of the the two-point function $\braket {F\O F}$ to the term $\Phi
\bbar \Phi$.

\EPSFIGURE[ht]{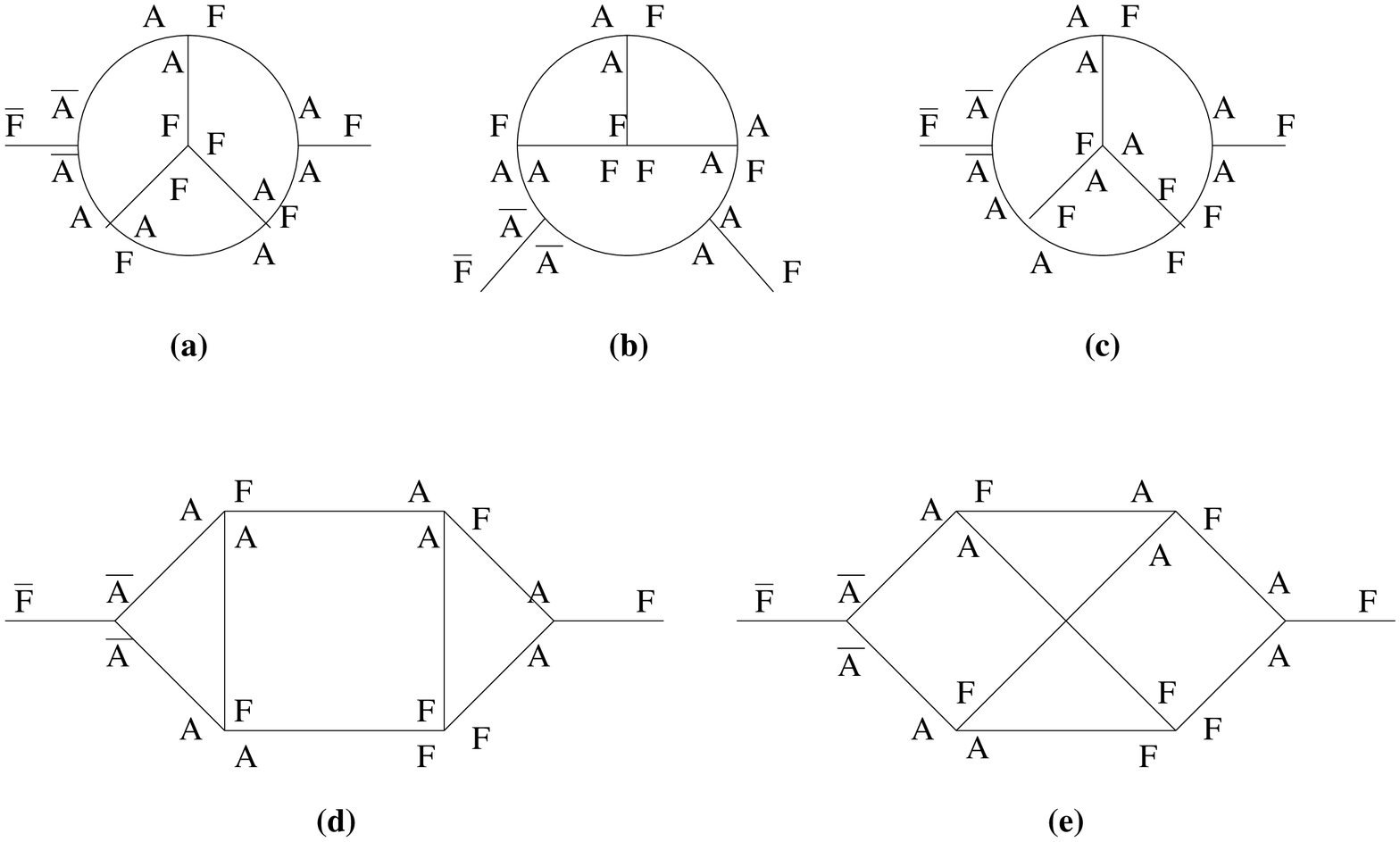,width=14cm}
{The three loop correction for $\braket {F\O F}$ with one $F^3$ insertion.
There are a lot of diagrams while we keep only five of them. The others
cancel between bosonic and fermionic loop contributions. For these
five remaining diagrams, there is no corresponding fermion loop contribution, and the result is finite.
\label{f:threeloop}
}

\section{Symmetry Considerations}
\subsection{Vanishing vacuum energy}

For the deformed Wess-Zumino model, we can identify two global $U(1)$
(pseudo)symmetries
by treating all coupling parameters, including $C^{\a\b}$, as the lowest components of (anti)chiral
superfields. They are $U(1)_\Phi$ flavor symmetry and $U(1)_R$
R-symmetry. Charge assignment is given as follows.

\begin{center}
\begin{tabular}{|c|c|c|c|c|c|c|c|} \hline
 & dim & $U(1)_R$ & $U(1)_\Phi$ &  & dim & $U(1)_R$ & $U(1)_\Phi$\\\hline
$\th$ & -1/2 & 1 & 0 & $\O \th$ &  -1/2 & -1 & 0 \\ \hline
$d\th$ & 1/2 & -1 & 0 & $d\O \th$ &  1/2 & 1 & 0 \\\hline
$A$ & 1 & 1 & 1 & $\O A$ & 1 & -1 & -1\\\hline
$\psi$ & 3/2 & 0 & 1 & $\O \psi$ & 3/2 & 0 & -1 \\\hline
$F$ & 2 & -1 & 1 & $\O F$ & 2 & 1 & -1 \\\hline
$g$ & 0 & -1 & -3 & $\O g$ & 0 & 1 & 3 \\\hline
$m$ & 1 & 0 & -2 & $\O m$ & 1 & 0 & 2 \\\hline
$C^{\alpha \beta}$ & -1 & 2 & 0 & $|C|$ & -2 & 4 & 0  \\\hline \hline
$\Phi$ & 1 & 1 & 1 & $\O \Phi$ & 1 & -1 & -1 \\\hline
\end{tabular}
\end{center}

Let $\Lambda$ be an ultraviolet cutoff scale. One can then
construct a set of couplings of mass-dimension $d$, charge $q_R=R$
and $q_\Phi=S$ as
\bea
\Lambda^d \O g^R \left({\O m\over \Lambda} \right)^{S-3R\over 2}
\left( g^4 ({\O m\over \Lambda})^6 |C| \Lambda^2\right)^Z
f(g\O g, {m\O m\over \Lambda^2}).
\label{couplings}
\eea
Here $f(x,y)$ is an arbitrary function of $x,y$, and $Z$ is a nonnegative integer. To show that the
vacuum energy is zero, we consider the vacuum-to-vacuum
amplitude directly:
\bea\label{partition}
\exp\Big(-\int \d^4 x E[C]\Big) =
{\cal Z}
=\int {\cal D}[\Phi] \exp\Big(-S_0+{g\over 3}  |C| \int \d^4 x
\Big[ (-{Q^2\over 4} \Phi)^2 \Phi \Big]_{\th^2} \Big). \eea
Here $S_0$ is the action of the ordinary Wess-Zumino model.
To compute quantum corrections due to the deformation, consider
the limit of small $C^{\a\b}$, and expand the energy density
perturbatively as
\bea E[C]= \Lambda^4 \sum_{n=1}^\infty |C|^n \left(g^4 \O m^6
\Lambda^{-4}\right)^n f_n \Big(g\O g, {m\O m\over \Lambda^2}\Big),
\label{vacuumenergy} \eea
where we have utilized the following observations: (1) the
$|C|$-independent contribution ($n=0$ term) is zero by the
standard non-renormalization theorem, and (2) the vacuum energy is
governed by the coupling Eq.(\ref{couplings}) with $d=4$ and
$R=S=0$.
Now take a partial derivative of equation (\ref{partition}) with respect to $|C|$, and then set $|C|=0$, to get
\bea
\langle {\rm vac} | -{g\over 3}  F^3 | {\rm vac}\rangle|_{C=0}
=  \Lambda^4\left(g^4 \O m^6 \Lambda^{-4}\right)^4
f_1(g\O g, {m\O m\over \Lambda^2}).
\eea
The left-hand side is the vacuum expectation value of the operator
${F^3}$ in the ordinary Wess-Zumino model.  However, since
$g F^3$ has charges $q_R=-4$ and $q_\Phi=0$, and
$U(1)_R$ is the global (pseudo)symmetry in the ordinary Wess-Zumino model,
the expectation value must be zero, so the function $f_1$ must vanish.
By taking partial derivatives successively and repeating the same sort
of argument, we can show that all of the $f_n$ vanish.
Thus the vacuum energy, as parametrized in equation (\ref{vacuumenergy}),
is zero.

\subsection{${\cal N}={1 \over 2}$ SUSY ground states}
In ordinary ${\cal N}=1$ supersymmetric theories, the
supersymmetric vacua are critical points of the holomorphic
superpotential $W(A)$ or, by conjugation, those of the
antiholomorphic superpotential $\O W(\O A)$. How about deformed,
${\cal N}={1 \over 2}$ supersymmetric theories?

We now argue that the perturbative supersymmetric vacua still
require
\beq\label{vac-def} \O W \, {}'(\O A) = 0. \eeq
As the antiholomorphic superpotential $\O W$ is not renormalized,
these vacua are stable against radiative corrections.

To show Eq.(\ref{vac-def}), it is sufficient to set all fermions
to zero and consider bosonic fields that are constant in space. In
this case, the tree-level part of the effective action takes the
form
\bea \Gamma = \int \d^4 x [F \O F + F W \, {}'(A) + \O F \O W \,
{}'(\O A) - \epsilon F^3], \nonumber \eea
where $\epsilon$ is an abbreviation for $-g |C|/3$. Notice that
the deformation term is proportional to $F$. Now consider
radiatively generated terms in the effective action. Since
$\partial_m = 0$ for constant bosonic fields, the operators
$Q_\a$, $D_\a$, and $\O D_\da$, are simply partial derivatives
with respect to $\th^\a$ or $\bth^\da$. So, after performing the
$\d^2 \th$ integral in the effective action, every radiatively
generated term must have at least one factor of $F$. Thus, the
full effective action for constant bosonic fields is expressible
in the form
\bea \label{lagforvac} \Gamma = \int \d^4 x \left[ F \O F + F W \,
{}'(A) + \O F \O W \, {}'(\O A) + F K(A,\O A, F, \O F) \right],
\eea
where $K(A,\O A, F, \O F)$ is a polynomial including both the
tree-level deformation term and the radiatively generated F- and
D-terms.

We now integrate out the auxiliary field $\O F$. Its equation of
motion from Eq.(\ref{lagforvac}) is
\bea 0 = F + \O W \, {}'(\O A) + F {\partial K \over
\partial \O F}. \nonumber \eea
We see immediately that $F$ is proportional to $\O W \, {}'(\O
A)$. Perturbatively, we can expand this proportionality factor
$1/(1+{\partial K \over \partial \O F})$ in powers of component
fields, and use this to replace $F$ in Eq.(\ref{lagforvac}). We
readily see that the effective action is proportional to $\O W'(\O
A)$. This means that, perturbatively, the scalar potential is of
the form
\beq V = \O W \, {}'(\O A) \Big[ W \, {}'(A) - H (A,\O A) \Big],
\label{scalarpotential} \eeq
where $H(A,\O A)$ denotes the aforementioned perturbation.
By solving ${\partial V\over \partial A}={\partial V\over \partial \O A}=0$, we find that a set of the ${\cal N}={1 \over 2}$
supersymmetric vacua with vanishing vacuum energy is given
precisely by the critical points Eq.(\ref{vac-def}) and $W
\,{}'(A) - H(A, \O A) = 0$.

Two remarks are in order. First, the scalar potential
Eq.(\ref{scalarpotential}) is in general complex-valued. This is
expected: the deformation has introduced non-Hermiticity to the
Lagrangian. Second, as $H(A, \O A)$ is renormalized at each order
in perturbation theory, solutions of $W \, {}'(A) - H(A, \O A) =
0$ are not stable under radiative corrections.

\subsection{${\cal N} = {1 \over 2}$ SUSY and antichiral rings}
The fact that the vacuum energy vanishes in the deformed Wess-Zumino
model leads to useful information concerning the vacuum state.
Recall that, after the deformation, the resulting ${\cal N}={1
\over 2}$ supersymmetry algebra is given by
\bea \left\{ Q_\a, Q_\b \right\} &=& 0 \nonumber \\
\left\{ Q_\a, \O Q_\da \right\} &=& 2 \sigma_{a \da}^m P_m
\nonumber
\\
\left\{\O Q_\da, \O Q_\db \right\} &=& 4 C^{(mn)}_{\da \db}P_m P_n
\label{1/2susyalgebra} \eea
where $C^{(mn)}_{\da\db} \equiv C^{\a\b} \sigma^m_{\a\da}
\sigma^n_{\b\db}$. Taking vacuum expectation values, we get a
nontrivial relation from the second line:
\bea 0 = \left< {\rm vac} | E | {\rm vac} \right> = \left< {\rm
vac} | Q_\a \O Q_\da + \O Q_\da Q_\a | {\rm vac} \right>, \nonumber
\eea
because the vacuum energy vanishes. Since $\O Q_\da$ corresponds
to the generator of explicitly broken supersymmetry (corresponding
to translation of $\bth$-coordinates), $\O Q_\da | {\rm vac}
\rangle$ does not vanish in general. Therefore, we are led to
conclude that
\bea Q_\a | {\rm vac} \rangle = 0 \quad \quad {\rm and} \quad
\quad \langle {\rm vac} | Q_\a = 0 \label{vacuumannihilation} \eea
for the ${\cal N}={1 \over 2}$ supersymmetric vacuum.

As the theory is defined by a non-Hermitian lagrangian, $|{\rm
vac} \rangle$ and $\langle {\rm vac} |$ are not a priori related,
but for ${\cal N}={1 \over 2}$ supersymmetric vacuum,  not only
$|{\rm vac} \rangle$ but also $\langle {\rm vac} |$ is annihilated
by $Q_\a$. Moreover, as the theory is defined on Euclidean space,
$Q_\a$ and $\O Q_\da$ are not hermitian conjugates, so the energy
of a given state is not necessarily positive-definite. Rather, as
discussed in the preceding subsection (see
Eq.(\ref{scalarpotential})), it is in general complex-valued.

Along with the non-renormalization of the antiholomorphic
superpotential, the relations Eq.(\ref{vacuumannihilation}) play
the crucial role for defining the antichiral ring. Recall that the
antichiral ring can be defined as a set of operators $\O {\cal O}$
obeying $\left[Q_\a, \O {\cal O} \right\} =0$. It then follows
that $\O {\cal O} \sim \O {\cal O} + \left[Q_\a, X \right\}$ for
every operator $X$, since
\bea \left< {\rm vac} \vert \left[Q_\a, X \right\} \O {\cal O}_2
\cdots \O {\cal O}_n | {\rm vac} \right>
&=&
\left< {\rm vac} |
Q_\a ( X \O {\cal O}_2 \cdots \O {\cal O}_n) | {\rm vac} \right> \pm
\left< {\rm vac} | X \left[Q_\a, \O {\cal O}_2 \right\}
\cdots \O {\cal O}_n | {\rm vac} \right> \pm \cdots \nonumber \\
&\pm& \left< {\rm vac} | X \O {\cal O}_2 \cdots \left[Q_\a, \O
{\cal O}_n \right\} | {\rm vac} \right> \pm \left< {\rm vac} | (X
\O {\cal O}_1 \cdots \O {\cal O}_n) Q_a | {\rm vac} \right>.
\nonumber \eea
From the definition of antichiral operators, all terms except the
first and the last ones vanish identically. It is here that the
conditions Eq.(\ref{vacuumannihilation}) come into play, ensuring
these remaining two terms vanish as well. Likewise, to demonstrate
that correlation functions of antichiral operators are independent
of separations and they factorize, one needs to show that
$\partial \O {\cal O}$ vanishes inside the correlation functions.
One again finds that, in proving this by using the
Eq.(\ref{1/2susyalgebra}) relation $\partial \O {\cal O} \sim
\left[Q, \left[\O Q, \O {\cal O} \right\} \right\}$, the
conditions Eq.(\ref{vacuumannihilation}) play a crucial role.

\section{Further Discussion}

In this work, we have studied quantum aspects of
${\cal N}={1 \over 2}$ supersymmetric field theories
in deformed superspace. We have
found many intriguing and surprising features that warrant 
further investigation. Here we discuss two points
we found interesting.

\hfill\break $\bullet$ The conclusions we have drawn in this work
are not from any specific choice of the superpotential. In fact,
for an arbitrary polynomial $W(\Phi)$ and $\O W(\O \Phi)$, one
readily finds that the Lagrangian may  still be recast as an
ordinary Wess-Zumino model with a number of deformation terms.
Explicitly, $\rm F$- and $\O {\rm F}$-terms are given in a compact
parametric form by
\bea L_F =&  {\cal K} \Big( F W'(A) - {1 \over 2} W''(A) \psi \psi
\Big), \qquad \qquad L_{\O F} = \Big(\O F \O W'(\O A) - {1 \over
2} \O W''(\O A) \O \psi \O \psi \Big). \nonumber \eea
where ${\cal K}$ is a functional differential operator:
\bea {\cal K} \equiv \int_0^1 \d \tau \cos \Big[ \tau |C|^{1/2} F
{\partial \over \partial A} \Big]. \nonumber \eea
Generalization to $N$-component Wess-Zumino model is
straightforward and replaces $F \partial_A$ by $F^a
\partial_{A_a}$.

The deformation part in the F-term contains operators of odd
powers of $F$ only, but, as evident from our analysis, operators
of all powers of $F$ are always generated radiatively. Large $N$
expansion of this model might find interesting applications to
various statistical mechanical systems.

\hfill\break

$\bullet$ The radiatively generated $F^2$-term is interesting. A
consequence would be that the term gives rise to mass splitting
between the boson and fermion component fields. To illustrate
this, consider the quadratic part of the Lagrangian, including an
$\epsilon F^2$-term, and integrate out the auxiliary fields, $F,\O
F$. The bosonic part of the Lagrangian becomes 
\bean
\O A\Box A - m
\O m A \O A +\epsilon \O m^2 \O A^2. 
\eean 
If we decompose field
$A$ into the real basis $A= a + i b$ we get 
\bean 
a \Box a+ b \Box
b- (|m|^2- \epsilon \O m^2) a^2 - (|m|^2+ \epsilon \O m^2) b^2 -
2i \epsilon \O m^2 a b, 
\eean
while the fermionic part of the
Lagrangian is unaffected. We see that the mass of the two real
bosons are split by $\pm \epsilon \O m^2$ and that the two real
bosons mix, analogous to the $K^0- \O K^0$ mesons, albeit the
theory is defined in Euclidean space.

The possibility that the deformation-induced $\epsilon F^2$ term
gives rise to mass splitting and flavor oscillation, while keeping
the vacuum energy to zero, might find interesting phenomenological
applications, once a suitable deformation can be achieved for
Lorentzian superspace.

\section*{Acknowledgements}
We are grateful to F. Cachazo, O. Lunin, and N. Seiberg for numerous
 helpful discussions. B.F. would like to thank V. Balasubramanian, A.
Naqvi, Y.H. He and V. Jejjala for discussions.
 R.B. and B.F. were supported by the NSF grant PHY-0070928.  S.J.R. was supported in part by the KOSEF Interdisciplinary Grant 98-07-02-07-01-5 and the KOSEF Leading Scientist Grant.  S.J.R. was a Member at the Institute for Advanced Study during this work.  He thanks the School of Natural Sciences for hospitality and for the grant in aid from the Fund for Natural Sciences.

\bibliographystyle{JHEP}

\end{document}